\documentstyle[aasms4,epsfig]{article}

\def\ie{{\it i.e.}~}
\def\etal{{\it et.al.}~}

\def\apj{{\it Ap.J.}~}
\def\mnras{{\it M.N.R.A.S.}~}
\def\ltsima{$\; \buildrel < \over \sim \;$}
\def\simlt{\lower.5ex\hbox{\ltsima}}
\def\gtsima{$\; \buildrel > \over \sim \;$}
\def\simgt{\lower.5ex\hbox{\gtsima}}

\begin{document}

\title{The Gravitational-wave contribution to the
CMB Anisotropies}

\author{Alessandro Melchiorri\altaffilmark{1}}
\affil{Dipartimento di Fisica, Universit\'a di Roma Tor Vergata}
\centerline {\it Via della Ricerca Scientifica, 00133 Roma, Italy}
\centerline{and}
\affil{D\'epartement de Physique Th\'eorique, Universit\'e de Gen\'eve}
\centerline {\it 24 Quai Ernest Ansermet, Ch-1211 Gen\'eve, Switzerland}

\author{Mikhail Vasil'evich Sazhin\altaffilmark{3}, Vladimir V. Shulga\altaffilmark{4}}
\affil{P. K. Sternberg Astronomical Institute}
\centerline {\it Universitetskii pr. 12, 119899 Moscow, Russia}

\author{Nicola Vittorio\altaffilmark{2}}
\affil{Dipartimento di Fisica, Universit\'a di Roma Tor Vergata}
\centerline {\it Via della Ricerca Scientifica, 00133 Roma, Italy}

\altaffiltext{1} {\it e-mail: melchiorri@roma2.infn.it}
\altaffiltext{2} {\it e-mail: vittorio@roma2.infn.it}
\altaffiltext{3} {\it e-mail: sazhin@sai.msu.su}
\altaffiltext{4} {\it e-mail: shulga@sai.msu.su}

\begin{abstract}

We study the possible contribution of a stochastic gravitational
wave background to the anisotropy of the cosmic microwave background
 (CMB) in cold and mixed dark matter (CDM and MDM) models.
We test this contribution against recent detections of CMB anisotropy
at large and intermediate angular scales. Our likelihood analysis
indicates that models with blue power spectra ($n \simeq 1.2$) and a
tensor to the scalar quadrupole ratio of $R = C_2^T / C_2^S \sim 2$
are most consistent with the anisotropy data considered here.
We find that by including the possibility of such background in CMB data
analysis it can drastically alter the conclusion on the remaining
cosmological parameters.
\end{abstract}


\section{ Introduction}

Inflationary theory has had a large impact on cosmology.
On the one hand, it resolves some difficulties of the standard Big-Bang model.
On the other, it provides a way of producing those density
 fluctuations that in
the gravitational instability scenario are the seed of the large scale structure
of the universe.  In fact, one of the most reliable predictions of the
inflationary paradigm is the parallel production of scalar and tensor
perturbations from quantum fluctuations of the inflaton field $
\hat{\phi} $ (\cite{sta1}; \cite{ruba}; \cite{sta}; \cite{abbo}).
The  amplitude  of tensor fluctuations
determines the value of the inflationary potential and, together with
other inflationary parameters, its first two derivatives
 (see {\it e.g.} \cite{turner}).
Thus, a detection of a nearly scale-invariant stochastic gravitational wave
(GW) background (tensor modes) is crucial in order to confirm any inflationary
model and constrain the physics occurring near the Planck scale, at
$\sim 10^{16}GeV$.

Observations of Cosmic Microwave
Background  (CMB) anisotropy promise to be unique in this respect
(\cite{staro}; \cite{crit}; \cite{turne}).
Recent numerical simulations (\cite{zalda}; \cite{dode};
 \cite{bond}) have shown that inflationary parameters will be measured
with an accuracy of  few percent by the MAP (\cite{map}) and Planck (\cite{planck})
space missions, which will
image the CMB anisotropy pattern with high sensitivity and at high angular
 resolution.

Meanwhile, the number of experiments reporting detections of
 anisotropy has increased to a couple of ten (see Table 1, below).
At the moment, the detections available seem compatible with
the predictions of inflationary models, like Cold Dark  Matter (CDM), with
 "blue" power spectra, i.e. $P(k)=Ak^{n_S}$ with
$n_S \simgt 1$ (\cite{deb};  \cite{benn}; \cite{bond2}).
As noticed by many authors,
there is a substantial rise in the anisotropy angular power
spectrum at $\ell \sim 200$,
which appears to be consistent with
the expected location of the first Doppler peak in flat models.
This small scale behaviour seems to disfavor a GW contribution.
In fact, as is well known, tensor fluctuations induce anisotropy only
on large angular scales ($ \ell \simlt 30$).
If there is a sizable contribution from GW in large scale detected
anisotropies, this would lower the predicted value of $(\Delta T / T)_{rms}$
on smaller scales.

Moreover, inflationary models that predict
$n_S \simgt 1$ generally predict vanishingly small tensor fluctuations
(\cite{kolb}).

Based on these arguments, a lot of  recent CMB data analysis
(\cite{line}; \cite{han}) has not taken into account the possible
presence of a GW background, assuming its contribution to be negligible.

In our opinion, there are two points that can alter these conclusions:

- Tensor modes are compatible with the theory of linear adiabatic
perturbations of a homogeneous and isotropic universe.
Like scalar perturbations and in contrast with vector perturbations,
they can arise from
small deviations from the isotropic Friedmann universe near
the initial singularity.
So, CMB data should be analyzed without
any {\it a priori} assumptions: the presence or absence of a
tensor component
in models with $n_S \ge 1$ can be only tested by observation.

- Few variations in the still undetermined
cosmological parameters (like the baryonic abundance or the Hubble constant)
and inflationary parameters (like the spectral index $n_S$) can
counterbalance the effect of tensor modes, increasing
the predicted value of $(\Delta T / T)_{rms}$
on small scales.

Thus, in this {\it paper}, we will discuss what kind of constraint
 present CMB anisotropy data provide on the tensor contribution allowing
all the remaining parameters to vary freely in their acceptable ranges.
We will extend our previous CMB data
analysis (\cite{deb}),
by including new CMB detections, and by analyzing a larger set
of models.  We restrict ourselves to critical universes ($\Omega_{matter}=1$), as
a recent analysis of CMB anisotropies and galaxy surveys (\cite{gawiser}) has shown
that pure scalar Mixed Dark Matter (MDM) models are in good agreement with the data set.
We will address the importance of a cosmological constant, reported by \cite{riess}
and \cite{perl}, in 
a forthcoming paper.

Since we treat the GW contribution as a free parameter,
we will not test any specific inflationary model.
So, our approach will be mainly phenomenological:
we assume that GW are created in the early universe
by some process during or immediately after inflation, which we
do not want to specify any further here.
Nonetheless, as the amplitude of the GW spectrum provides a test for inflation
(see next section), in our conclusions
we will discuss if
 results are compatible with this paradigm.

Since any possible GW
signal will affect the matter power spectrum normalization
inferred from COBE, we will test the models
that best fit the CMB data with the
normalization $\sigma_8$ of the matter fluctuation in $8 h^{-1}$
spheres and with the shape of the spectrum
from the \cite{peaco} analysis.

The plan of the paper is as follows. In Sect.2 we write the set of
equations necessary to describe the inflationary process
in the slow roll approximation.  In Sect.3 we briefly discuss the analysis of
the current
degree-scale CMB experiments.
In Sect.4 we test the best fit models  with
  the Large-Scale Structure (LSS) data.
Finally, in Sect. 5 we present and discuss our conclusions.

\section{Early Universe}

Inflation in the early universe is determined by the potential $V(\hat
\phi)$, where $\hat \phi$ can be a multiplet of scalar fields.
Here we restrict ourselves to the case of a single, minimally coupled
scalar field $\phi$ with potential $V$ and equation of motion

\begin{equation} \ddot
\phi + 3 H \dot \phi + V'=0,
\end{equation}

\noindent (as usual,  the dot and prime indicate
derivatives with respect to physical time $t$ and to the scalar field
$\phi$, respectively). The expansion rate in the early universe can be
written as:
\begin{equation}
H^2={8\pi\over 3 m_{Pl}^2} \biggl[{1\over 2} {\dot\phi}^2 + V(\phi)\biggr]
\end{equation}
where $m_{pl}=1.2\cdot 10^{19}GeV$ is the Planck mass (we use natural
units, i.e. $h=c=k=1$).

The slow-roll approximation holds in most of the inflationary
models. This condition is valid if (\cite{cop}; \cite{hod})
\begin{equation}
{m_{pl}^2\over 4\pi}\biggl[ \frac{H''}{H}\biggr] = \eta (\phi) << 1
\end{equation}
and
\begin{equation}
{m_{pl}^2\over 4\pi} |\frac{H'}{H}|^2 = \epsilon (\phi) << 1
\end{equation}

The second condition, since $\epsilon$ is a direct measure of the equation of
state
of the scalar field matter, also implies the period of accelerated expansion
(\cite{dode}).

In the slow roll approximation the amplitude of scalar and tensor
perturbations are related to the
inflationary potential as follows (\cite{cop}):
\begin{equation}
A_S(\phi)= \sqrt{2\over \pi}{1\over
m_{pl}^2}\frac{H^2}{|H'|}
\end{equation}
\noindent and
\begin{equation}
A_T(\phi)= {1\over \sqrt{2}\pi }{H\over m_{pl}}
\end{equation}

We can relate the wavelength, $\lambda$,
and the Hubble parameter during inflation, $H(\phi)$,
with the scalar field by writing:

\begin{equation}
{d\ln \lambda \over
  d\phi}={\sqrt{4\pi} \over m_{pl}}
\frac{A_S}{A_T}
\end{equation}

\noindent and

\begin{equation}
{\partial \ln H \over  \partial\phi}={\sqrt{4\pi} \over m_{pl}}
\frac{A_T}{A_S},
\end{equation}

\noindent respectively.

Let us define the spectral equations for scalar and  tensor
components as follows:

\begin{equation}
A_S^2(k)=A^2 ({k\over k_0})^{\displaystyle n_S-1}
\end{equation}

\noindent and

\begin{equation}
A_T^2(k)=B^2({k\over k_0})^{\displaystyle n_T}
\end{equation}

 \noindent where $k_0=H_0$ is the wavenumber of a fluctuation which re-enters
 the horizon
at the present time, and $A$ and $B$ are constants.  It is easy to see that
$n_T=-2\epsilon(k)$  if $\lambda=\lambda_0$, and $n_T=0$ if $\lambda 
\rightarrow 0$
(\cite{lid}).

We define  the ratio of amplitudes of the scalar and tensor
 modes by:

\begin{equation}
r = \sqrt{\epsilon (k_0)} ={\displaystyle B\over \displaystyle A}
\end{equation}

By solving Eq.(7) and assuming $n_T \sim 0$, the scalar field can be written
as a function of the wavelength:

\begin{equation}
\phi(\lambda)=\phi_0 +\phi_1 \biggl[
(\frac{\displaystyle \lambda}{\displaystyle \lambda_0})^{\displaystyle
n_S-1 \over \displaystyle 2} - 1\biggr]
\end{equation}

\noindent where $\phi_0$ is a constant, to be found from boundary conditions,
 and $\phi_1=\frac{\displaystyle
r}{\displaystyle n_S - 1}\frac{\displaystyle m_{pl}}{\displaystyle
\sqrt{\pi}}$.

 Furthermore,  Eq.s (12) and (8) allow us to
  find the Hubble parameter $H(\phi)$ during inflation:

\begin{equation}
H(\phi)=H_i \exp(\frac{\displaystyle r^2}
{\displaystyle n_S-1}\displaystyle \xi^2)
\end{equation}

\noindent where $\xi=\frac{\displaystyle \phi +\phi_1 -\phi_0}{\displaystyle
\phi_1}$ and $H_i$ is a constant.

The potential can be written in terms of the Hubble parameter:

\begin{equation}
V(\phi)=\frac{\displaystyle 3 m_{pl}^2}{8\pi}
H^2(\phi)
\end{equation}

At this point, we can  define the relation
between the
quadrupole multipoles of the CMB anisotropy generated by
scalar and tensor perturbations: $C_2^S$ and $C_2^T$,
respectively.
To do this we will follow the calculations done
by (\cite{sour}) in which both $C_2^S$ and $C_2^T$ were found as a function of
$H(\phi)$ at $\lambda=H_0^{-1}$. So we have:

\begin{eqnarray}
C_2^S=\frac{2\pi^2}{25} f(n_S) \frac{1}{m_{pl}^4}
\frac{H^4}{(H')^2} \\
C_2^T= \frac{2.9}{5\pi} \frac{H^2}{m_{pl}^2}
\end{eqnarray}

\noindent and

\begin{equation}
\frac{C_2^T}{C_2^S}=\frac{29}{4\pi^3}  \frac{m_{pl}^2}{f(n_S)}
\frac{(H')^2}{H^2}
\end{equation}

\noindent where

\begin{equation}
f(n_S)=\frac{\displaystyle \Gamma(3-n_S)\Gamma({3+n_S \over
2})}
{\displaystyle\Gamma^2({4-n_S  \over 2})\Gamma({9-n_S \over
2})}
\end{equation}

Using Eq.(13) we can  write:

\begin{equation}
\frac{\partial \ln H(\phi)}{ \partial \phi}=\frac{2 \sqrt{\pi}
}{m_{pl}} r \xi
\end{equation}

Therefore, at $\phi=\phi_0$  (\ie  $\xi=1$),
we have:

\begin{equation}
 R(n_S) \equiv
\frac{C_2^T}{C_2^S} = {{29 r^2} \over{ \pi^2 f(n_S)}}
\end{equation}

\noindent As we can see from the equation above, the
tensor to scalar quadrupole ratio $R$ is related to the slow-roll
parameter $\epsilon$. Eq. (20) identifies a region in the
($n_S$,$R$) space of values where the slow-roll condition is
satisfied. Furthermore, as $\epsilon < 1$ only if the
universe has undergone a period of accelerated expansion, one can use
this equation to test the inflationary scenario.

In the same way, we can use Eq.(19) in Eq.(3) in order to find:

\begin{equation}
2\eta=n_s-1+2r^2
\end{equation}

\noindent so the slow roll condition (4) implies Eq. (3) if $n_s \sim 1$.

Using Eq.(14), we can now write the potential as

\begin{equation}
V(\phi_0) = {{15} \over {23.2}} C_2^T m_{pl}^4
\end{equation}

\noindent Therefore, the measurement of the contribution to the quadrupole
 anisotropy
 of tensor fluctuation, $C_2^T$,
allows us to estimate the size of the potential responsible for inflation.

\section{ CMB Anisotropy}

\subsection{Method}

We use a set of the most recent CMB anisotropy detections,
both on large and degree angular scales, in order to
estimate the amplitude of tensor fluctuations.
The likelihood of the assumed
independent CMB anisotropy data is (see \cite{deb}):

\begin{eqnarray}
{\cal L} & = & \prod_j {1 \over {(2 \pi [
({\Sigma_j}^{(the)})^2 + ({\Sigma_j^{(exp)}})^2])^{1/2}}} \nonumber \\
&\times & exp (- \frac {1} {2} \frac{[{\Delta_j}^{(exp)} -
\Delta_j^{(the)}]^2}{(\Sigma_j^{( the)} )^2
+ ({\Sigma_j^{(exp)}})^2} )
\end{eqnarray}

\noindent where $\Delta_j^{exp}$ and $\Delta_j^{the}$
are the experimentally
detected and theoretically expected mean square anisotropy,
respectively. The $(\Sigma_j^{(the)})^2$
and $(\Sigma_j^{(the)})^2$
are the respective cosmic and experimental variances.
Obviously, the likelihood depends on
the parameters of the cosmological model.
Although a complete analysis should cover all the parameter space,
here we restrict ourselves to flat
models ($\Omega_0 = 1$) composed of baryons
($0.01 \simlt \Omega_b \simlt 0.14$), cold
dark matter ($\Omega_{CDM} \simgt 0.7$), hot dark
matter ($\Omega_{\nu} \simlt 0.3$), photons and
massless neutrinos.
As shown in (\cite{deb}; \cite{chun}; \cite{dode2}) the angular power spectrum
 of MDM models differs from pure CDM by less than $10\%$
  in the angular scales of interest.
Given the poor sensitivity of the available CMB
anisotropy detections at degree angular scales,
we restrict ourselves to  pure CDM models, keeping
in mind that basically the same power spectrum is
also expected for MDM models. The predictions
of CDM and MDM models for the matter power spectrum obviously differ, and in
a substantial way:
we will discuss this point in more detail below.

Here we keep as free parameters $\Omega_b$ and
$h$. Both  parameters affect the positions and amplitudes
of the so-called Doppler peaks of the angular power
spectrum. In fact, changing $\Omega_b$ at fixed $h$
changes the pressure of the baryon-photon fluid before
recombination, increasing its oscillations below its
Jeans length.  A larger baryon to photon
ratio will increase the compressions (which produce the even peaks
in $C_\ell$ for inflationary models) and
decrease the rarefaction (odd peaks for inflationary models).
Lowering $h$ at fixed $\Omega_b$
 changes the epoch of
matter-radiation equality:  potentials inside the horizon
decay in a radiation dominated era but not in a fully matter dominated one.
The combination $\Omega_bh^2$, which actually appears
in the calculations, is also constrained by primordial
nucleosynthesis arguments (\cite{copi}):
$ 0.01 \simlt \Omega_bh^2 \simlt 0.026$.
Moreover, from globular cluster ages, $0.4 \simlt h \simlt 0.65$
(\cite{kotu}).

We will also explore variations in the spectral index of the
(scalar) primordial power spectrum $n_S$.
We restrict ourselves to values of $n_S \simlt 1.5$, to
be consistent with the absence of spectral distortions
in the COBE/FIRAS data (\cite{hu2}).
A parameter independent normalization for the power
spectrum can be expressed in terms of the amplitude of the
multipole $C_{10}$.
We define the parameter ${\cal A} \equiv A/A_{COBE}$ as
the amplitude ${\cal A}$ of the
power spectrum (considered as a free parameter) in units of $A_{COBE}$, the
amplitude needed to reproduce
$ C_{10} \sim 47.6 {\mu K}^2 $, as observed on the COBE-DMR
four-year maps (\cite{bunn}).

Finally, for tensor fluctuations, we will assume $n_T = 0$.
In fact, variations in the tensor spectral index in the
range $-1 \simlt n_T \simlt 0$ do not give appreciable
changes in the structure of the $C_\ell$'s, given the cosmic variance
and the current experimental sensitivity.
We parameterize the amplitude of these tensor fluctuations
with $R$, defined in Eq. (22).
So, in the end, we will consider only five
quantities as free parameters:
 ${\cal A}$, $n_S$, $R$, $\Omega_b$ and $h$.

We have computed the angular power spectrum of CMB anisotropy
by solving the Boltzmann equation for fluctuations
in CMB brightness (\cite{yu}; \cite{hu}).
 Our code is described in (\cite{deb}; \cite{melvit})
 and allows the study of CMB anisotropy both in cold (CDM) and
mixed (MDM) dark matter models.  Our $C_\ell$'s match
to better than  $0.5 \%$ for $\ell \le 1500$ compared with those
of other codes (\cite{seljak}; \cite{chun}).
In Fig. 1 we show the $C_\ell$'s for different
parameter choices.

The data we consider are listed in Table 1 and shown in
Fig. 1.
We have updated the data presented in our previous paper
(\cite{deb}) to
include the new results from the Tenerife, MSAM and CAT experiments.
For the COBE data, we use the $8$ data points from
\cite{tegma}, that have the advantage of uncorrelated
error bars.

\subsection{ Results}

The best fit parameters (i.e. those which
maximize the likelihood) are (with $ 95 \%$ confidence):
$n_S=1.23^{+0.17}_{-0.15}$, $R=2.4^{+3.4}_{-2.2}$, with ${\cal A} = 0.92$,
$\Omega_b = 0.07$ and $h= 0.46$. We can only put the following
upper limits (at $68 \%$) on these last two best fit values: $\Omega_b <0.11$,
$h < 0.58$.

A probability confidence
level contour in the five dimension volume of parameters
is obtained by cutting the
${\cal L}$ distribution with the isosurface ${\cal L}_P$,
 and by requiring that the
volume inside ${\cal L}_P$ is a a fraction $P$ of the total
volume. The projections of the ${\cal L}_{68}$ and ${\cal L}_{95}$
 surfaces on the $n_S-R$ plane are shown in Fig. 2.

As we can see from Fig. 2, the Likelihood contours are very
broad and models with spectral index $n_S \sim 1$ and $R=0$ are
statistically indistinguishable from models with $n_S \sim 1.4$ and $R
\sim 4$.

The quite large values of  $R$ for
$n_S \simgt 1 $ are due
to a parameter degeneracy problem that present CMB anisotropy  detections
are not able to solve (see Fig. 1).  In fact, increasing the contribution
 of tensor modes
 boosts the anisotropy
on large scales ($>>2^\circ \Omega_0^{1/2}$).  As the theoretical predictions
 are normalized
to COBE/DMR, adding tensor fluctuations while keeping all the other
 parameters fixed,
actually suppresses  the level of degree scale anisotropy. To counterbalance
 this effect, it is
necessary to postulate "blue" primordial spectra, \ie $n_s\simgt 1$.
 The shape of the
confidence level region in the $n_s-R$ plane reflects this correlation.
 This degeneracy in the
model prediction is actually broken at a higher angular resolution,
$\ell \simgt 300$ say, where present
experiments are particularly affected by cosmic variance,
due to the very small region of the sky sampled (see Table 1).
We have the following $95 \%$ C.L.
upper limits on $R$: $0.3$, $1.3$,
$2.5$, $4.5$, $7.8$ and $12.5$
for $n_s =0.8$, $0.9$, $1.0$, $1.1$, $1.2$ and $1.3$,
 respectively. At $n_s =1.4$ and $1.5$ we can put
$95 \%$ C.L. lower limits of $1.0$ and $2.8$ on $R$.
A quadratic fit to the maxima
distribution gives:

\begin{equation}
R=34.3-70.8n_S+36.5n_S^2
\end{equation}

\noindent for $1.1 \le n_S \le 1.5$.
With the above equation,
we find that the tensor component can have an
rms amplitude value of $\sim 28 \mu k$ for $n_s=1.1$ and
 $\sim 49 \mu k$ for $n_s =1.5$, while the
scalar component remains at $\sim 100 \mu k$.

\

It is interesting to see (Fig.1) that models with $n_S \sim 1.4$ and
$R \sim 3$, which are well compatible with our analysis,
seem to prefer a greater Hubble constant, $h \sim 0.6$.
So, the gravitational wave contribution also seem to
moderate the discrepancy between the value
$h \sim 0.7$ (\cite{free}), inferred by
several different methods, and the value $h \sim 0.4$ (\cite{line})
inferred by scalar-only CMB analysis.

\

We found that inside the $95 \%$ contour, the overall
normalization amplitude, in units of $A_{COBE}$ is ${\cal A} = 1 \pm 0.2$,
{\it i.e.} all the models considered therein correspond well  with
 COBE/DMR normalization.

\

The simple analysis carried out here does not take
into account the correlation due to overlapping sky
coverage (e.g., Tenerife and COBE, and/or MSAM and
Saskatoon). We check the stability of our analysis
with a jacknife test, i.e. removing one set of experimental
data each time. We have a maximum variation of $ 3 - 4 \%$ in our limits
in the $n_S-R$ plane, except with the removal of COBE data that
 modifies our results by $\sim 10 \%$.
So, neglecting this correlation does not significantly change
the results of our analysis.
We also repeated the analysis including the possible $\pm 14 \%$
calibration error to the five Saskatoon points (\cite{net}),
and we did not find significant variations.
In the limited cases where comparison is possible,
our analysis produced results similar
to those of \cite{bond2}, \cite{line} and \cite{han}.

\section{ Comparison with LSS}

As we have seen, blue models with a substantial tensor
component agree well with CMB data.
Tensor modes have dramatic effects on the matter
power spectrum, reducing its normalization by a factor of
$(1+R)^{-1}$.
Using the above fit formula, the tensor contribution to the
CMB correlation function on the COBE/DMR scales
can be between $54 \%$ for $n_s=1.1$ and $91 \%$ for $n_s=1.5$.
In this section we want to test these models
with large scale matter distribution.
As is well known, CDM blue models
predict a universe that is too inhomogeneous on scales
$\le 10 h^{-1}Mpc$.
Nonetheless, the excess power on these scales can be reduced by considering a
mixture of cold and hot dark matter, i.e. mixed dark matter (MDM)
models. The difference in the $C_l$ behaviour between a pure CDM and
an MDM ($\Omega_\nu \le 0.3$) model is very tiny, $\le 2 \% $ up to
$l \sim 300$ and $\le 8 \%$ up to $l \sim 800$ (see, for example \cite{dega}).
Therefore, the results of our CMB analysis are the same in this kind of model.
In Figure 3, MDM matter power spectra
from models that agree with CMB data are shown.
The data points are an estimate of the linear power spectrum from
\cite{peaco}, assuming a CDM flat universe and bias values
between Abell, radio, optical, and IRAS catalogs $b_A:b_R:b_O:b_I=
4.5:1.9:1.3:1.0$ with $b_I=1.0$.
As shown in (\cite{clay}) recovered linear power spectra of
CDM and MDM models are nearly the same in the region $0.01 \le k \le 0.15
 h Mpc^{-1}$ but diverge from this spectrum at higher $k$, so
we restrict ourselves to this range.
The $\chi^2$ (with $11$ degrees of freedom) are $15$, $10$,
$21$, $9$, $37$, $53$ for models in Figure 3 with $(1.4,3.3)$, $(1.3,3.9)$,
$(1.2, 1.3)$, $(1.1,0.6)$,$(1.0,0.1)$ and $(0.9, 0)$ in the $(n_s,R)$ space.
So, models with a large tensor
contribution on COBE scales and blue spectral index seem to agree well
also with the shape of matter distribution on large scale.
The values for the $\sigma_{8}$, computed with CMBFAST (\cite{selja}),
 are $0.69$, $0.61$, $0.66$, $0.63$, $0.63$, $0.74$, 
in very reasonable agreement with the value of 
$\sigma_8^{IRAS} = 0.69 \pm 0.05$ (\cite{fisher}) derived from
the IRAS catalog.

Whether IRAS galaxies are biased is still under debate.  
Analysis from cluster data (\cite{eke}, \cite{pen}, \cite{bryan}), shows
a preferred value of $\sigma_8 \sim 0.5 - 0.6$ 
with few percent error bars.
 Analysis from peculiar velocities (\cite{zehavi}) 
results in a larger value
$\sigma_8 = 0.85 \pm 0.2$, which seems to be in severe conflict 
with the cluster data. Thus, the theoretical values of $\sigma_8$ for
blue MDM models with a relic gravitational wave background are
between the $\sigma_8$ values derived from cluster abundance and
peculiar velocities. In any case, the likelihood of the CMB data is
quite flat around its maximum. So, it is easy to find models,
 statistically indistinguishable from the best fit models,
with $\sigma_8$ nearer either to $0.5$ or to $0.8$.

Because of statistical and/or systematic uncertainties
we do not consider it appropriate to put more than qualitative
conclusions on these results, but still one can say that
the lower matter normalization due to the
tensor component helps the blue MDM models to match the LSS data.

\section{ Conclusions}
\label{sec:level5}

Our main conclusions, are as follows:

1. The conditional Likelihood shows a maximum at  $n_S=1.23^{+0.17}_{-0.15}$,
 $R=2.4^{+3.4}_{-2.2}$, with ${\cal A} = 0.92$,
 $\Omega_b = 0.07$ and $h= 0.46$. Thus,
 there is some evidence that a tensor component can be
present, and in a substantial way, in models with $n_s$ greater
than one.
Inflationary models of this type have been investigated
by \cite{cop} and by \cite{luk} and thus belong to the class of hybrid 
inflationary models (\cite{kin}). The general form of the 
potential can be written as $V(\phi)=V_0 + {1 \over 2 } \mu^2 \phi^2$
At the end of inflation, the inflationary potential $V(\phi)$ is not equal to
zero, being $V_0$ of the order of $ \sim (6 10^{16} Gev)^4$.
 In order to be consistent
with the present vacuum energy $\le (10^{-30} Gev)^4$, one additional
field is necessary to finish inflation. The inclusion of this
field does not change the conclusion of our analysis, 
since it affects only the high frequency region of the GW spectrum 
($\sim 100 MHz$).
For models on the best fit curve (Eq.24), $V(\phi_0)$ belongs to the
interval $4.3 \cdot 10^{-11} m_{pl}^4 < V_0 < 1.3 \cdot 10^{-10}
m_{pl}^4$.
In Fig.2 we plot Eq.(24) with the condition $\epsilon=1$.
The region below this curve in the $n_S - R$ plane
is where the slow roll approximation is valid.
As we can see, models on our best fit curve satisfy this condition,
even if models with $\epsilon \ge 1$ are compatible with observations.

Approaching the limiting region $\epsilon =1$, higher order terms in the
slow roll approximation became valuable.  This leads to changes
in our conclusions on the potential by a factor $1-\epsilon / 3 \sim 30 \%$ 
(\cite{kin}).

2. The $95 \%$ region on the $n_S-R$ plane includes a wide range of
parameters. This means that the presently available data set is
not sensitive enough to produce precise determinations for $n_S$
and $R$.
Systematic and statistical errors in the different experiments
are still significant, but, as we have shown,
the difficulties involved in such determinations are
 mainly due to a degeneracy in these parameters.
So, the $(n_S,R)$ degeneracy has important consequences for
tests of the inflationary theory:
increasing the scalar spectral index and the
tensor component lead to a break in the slow roll approximation, but it
also produces CMB power spectra near to the scale invariant one.
Therefore it is difficult from the present CMB data to see if the slow roll
condition is correct.

Furthermore, current CMB results on the normalization of the
matter power spectrum and/or its spectral index can
be biassed and/or anti-biassed by a huge tensor contribution.
As we can see from Fig.1, this degeneration also has effects
on the constraints of the remaining cosmological parameters,
being a model with $h \sim 0.6$ statistically indistinguishable
from a model with $h \sim 0.4$.

The inflationary background of primordial
gravitational waves is assumed detectable mainly through CMB
experiments. The local energy density of this background is, in the most
optimistic situation, extremely low, with $d \Omega_{GW} h^2 / d log k \sim
10^{-16}$ at frequencies $10^{-15} {Hz} < f < 10^{15} {Hz}$.
The tenuity of this signal makes the degeneracy in the $n_S$ and $R$
parameters much more worrying than similar degeneracy in other
parameters (e.g. $h$ and  $\Omega_b$) that could be constrained through
other measurements.

3. "Blue" MDM models with a tensor contribution, are
in reasonable agreement with the present values of $\sigma_8$, and with
the shape of the matter power spectrum inferred by the \cite{peaco}
analysis. A tensor contribution could also be a viable mechanism in
order to reconcile these models with a low value for the
$\sigma_8$ around $\sim 0.5$ (\cite{henry}).

This being the situation, a measure of the structure of the secondary peaks
becomes a crucial test for the presence of tensor perturbations.
Using the above best fit equation,
we can make some predictions regarding future detections.
We found that an experiment
with a window function probing the multipoles $500 \le \ell \le 680$,
will measure a total rms anisotropy of $28.3 \mu K$ for $n_s=1.1$, and
 $34.4 \mu  K$,
with $n_S=1.5$.
An $\sim 20 \%$ difference that could be proven, when
the sensitivity of these experiments is within a few $\mu k$,
with an improved sky coverage.
Polarization measurements at intermediate angular
scales can also be helpful (\cite{sazh}, \cite{poln}, \cite{sazb},
\cite{sazhshul}).  The possibility of a direct separation of scalar
perturbations from tensor perturbations by the method of decomposition
of Stocks parameters in sets of spin $\pm 2$ spherical harmonics seems
extremely promising (\cite{kam}, \cite{sel}, \cite{sazsh}).

Possibly a definitive answer will come
when future CMB experiment provides a clear and robust
picture of sub-degree angular scale anisotropy and polarization.

\acknowledgments

We wish to thank Paolo de Bernardis, Ruth Durrer, Giancarlo De Gasperis, 
Martin Kunz
 and Andrew Yates. M.V.S. acknowledges the University of "Tor Vergata" for 
hospitality during writing part of this paper. MVS 
acknowledge "Cariplo Foundation" for Scientific Research and 
"Landau-Network - Centro Volta"  for financial support during the writing 
the last version of this paper.

\newpage

\begin{figure}
\caption{Power spectra of CMB anisotropies for different combinations
of inflationary and cosmological parameters. The data points are derived
from the experiments listed in Table 1.}
\end{figure}

\begin{figure}
\caption{ Confidence level ($68$ and $95 \%$) regions for the spectral
index $n_S$ and the tensor to scalar quadrupole ratio $R=C_2^T/C_2^S$.
The region below the black curve is where the slow-roll approximation
is valid.}
\end{figure}

\begin{figure}
\caption{ Matter power spectra for MDM models.
All models are normalized to the 4-year COBE/DMR data
using the method of Bunn and White 1997.}
\end{figure}


\newpage
\begin{deluxetable}{crrrrrrr}
\footnotesize
\tablecaption{CMB Anisotropy detections used in the Analysis. References are:
(0) Tegmark and Hamilton 1997; (1) Hancock et al. 1994;
(2)  Gutierrez et al. 1997;
(3) Gundersen et al. 1993;
(4) Dragovan et al. 1993; (5) de Bernardis et al. (1994); (6) Masi et al 1996;
(7) Tanaka et al. 1996; (8) Cheng et al. 1994; (9) Cheng et al. 1996;
(10) Cheng et al. 1997; (11) Netterfield et al. 1996; (12) Scott et al. 1996;
(13) Baker et al. 1997.
\label{tbl-1}}
\tablewidth{0pt}
\tablehead{
\colhead{Experiment}&\colhead{Reference}&\colhead{$\Delta T^2 (\mu K^2)$}&\colhead{ $\ge 68 \% (\mu K^2)$}
&\colhead{$\le 68 \% (\mu K^2)$}&\colhead{Sky Coverage}
&\colhead{${\ell}_{eff}$}}

\startdata
COBE1 & 0& 25.2 & 183 & 25.2 & 0.65 & 2.5\nl
COBE2 & 0& 212 & 126 & 128 & 0.65 & 3.3\nl
COBE3 & 0& 256 & 96.5& 96.9& 0.65& 4.1\nl
COBE4 & 0& 105.5 & 48.3&48.2 & 0.65& 5.5\nl
COBE5 & 0& 101.9 & 26.5& 26.4& 0.65& 8.1\nl
COBE6 & 0& 63.4 & 19.11& 18.9& 0.65& 11.6\nl
COBE7 & 0& 39.6 & 14.5 & 14.5& 0.65& 16.7\nl
COBE8 & 0& 42.5 &12.7& 12.8& 0.65& 25.1\nl
Tenerife & 1& 1770 & 840 & 670  & 0.0124& 20.1\nl
Tenerife & 2& 3975 & 2855 & 1807 & 0.0124& 20.1\nl
South Pole Q& 3& 480& 470& 160 & 0.005& 49.4\nl
South Pole K& 3& 2040& 2330& 790 & 0.005& 65.7\nl
Python& 4 & 1940& 189& 490& 0.0006& 129.0 \nl
ARGO Hercules& 5& 360& 170& 140& 0.0024&118.9\nl
ARGO Aries& 6& 580& 150& 130& 0.0024&118.9\nl
MAX HR& 7& 2430& 1850&1020&0.0002&162.0\nl
MAX PH& 7& 5960& 5080&2190&0.0002&162.0\nl
MAX GUM& 7& 6580& 4450&2320&0.0002&162.0\nl
MAX ID& 7& 4960& 5690&2330&0.0002&162.0\nl
MAX SH& 7& 5740& 6280&2900&0.0002&162.0\nl
MSAM93 & 8&4680&4200&2450&0.0007&179\nl
MSAM94 & 9& 4261 & 4091 & 2087  & 0.0007& 179\nl
MSAM94 & 9& 1960 & 1352 &  858  & 0.0007& 251\nl
MSAM95 & 10& 8698 & 6457 & 3406  & 0.0007& 179\nl
MSAM95 & 10& 5177 & 3264 & 1864  & 0.0007& 251\nl
Saskatoon&11&1990&950&630&0.0037&99.9\nl
Saskatoon&11&4490&1690 & 1360&0.0037&175.4\nl
Saskatoon&11&6930&2770&2140&0.0037&235.2\nl
Saskatoon&11&6980&3030&2310&0.0037&283.2\nl
Saskatoon&11&4730&3380&3190&0.0037&347.8\nl
CAT1 & 12& 1180 & 720 & 520  & 0.0001& 414\nl
CAT2 & 12& 760 & 760 & 360  & 0.0001& 579\nl
CAT1 & 13& 934 & 403 & 232  & 0.0001& 414\nl
CAT2 & 13& 577 & 416 & 238  & 0.0001& 579\nl

\enddata
\end{deluxetable}
\clearpage

\plotone{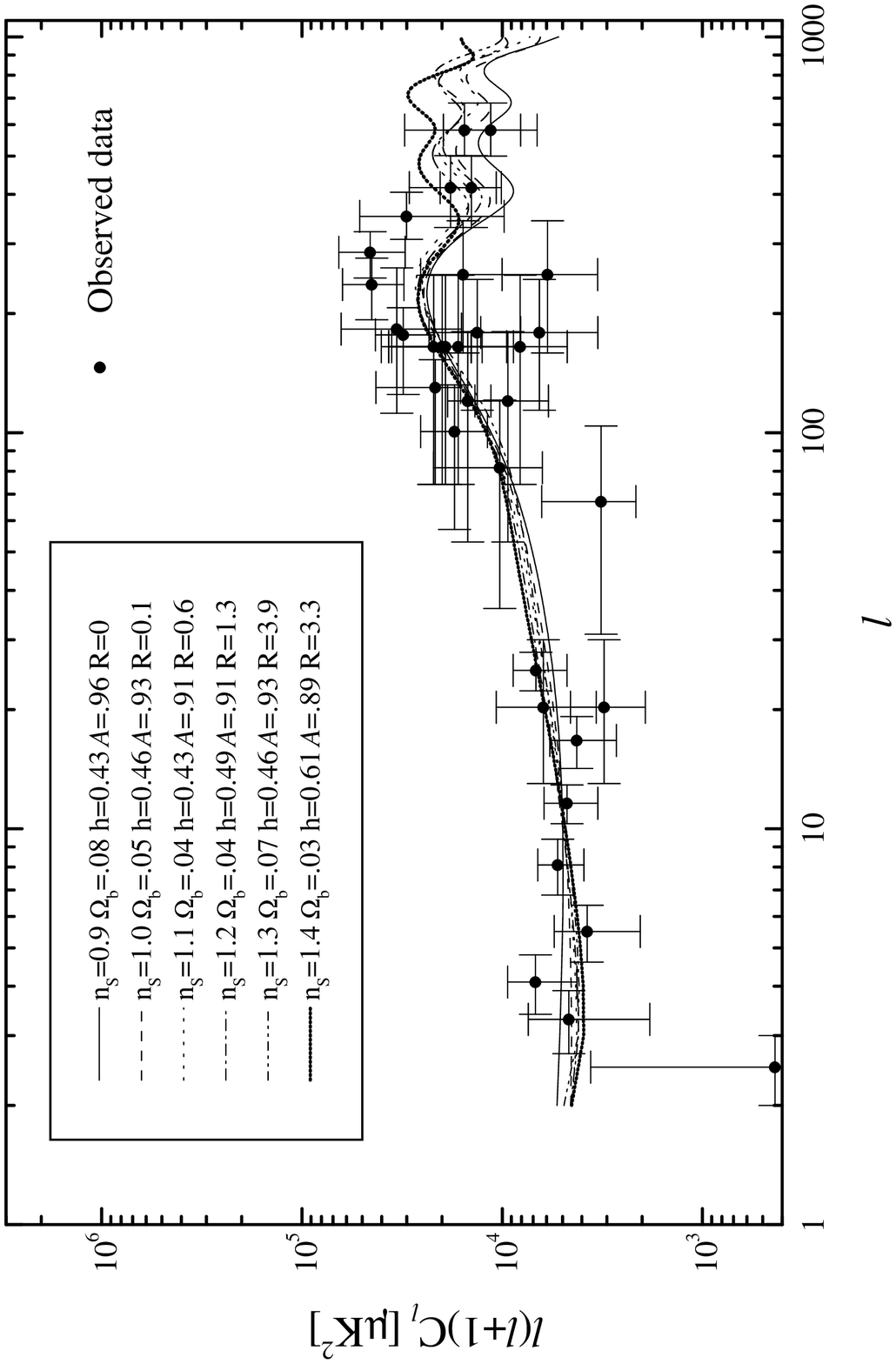}

\plotone{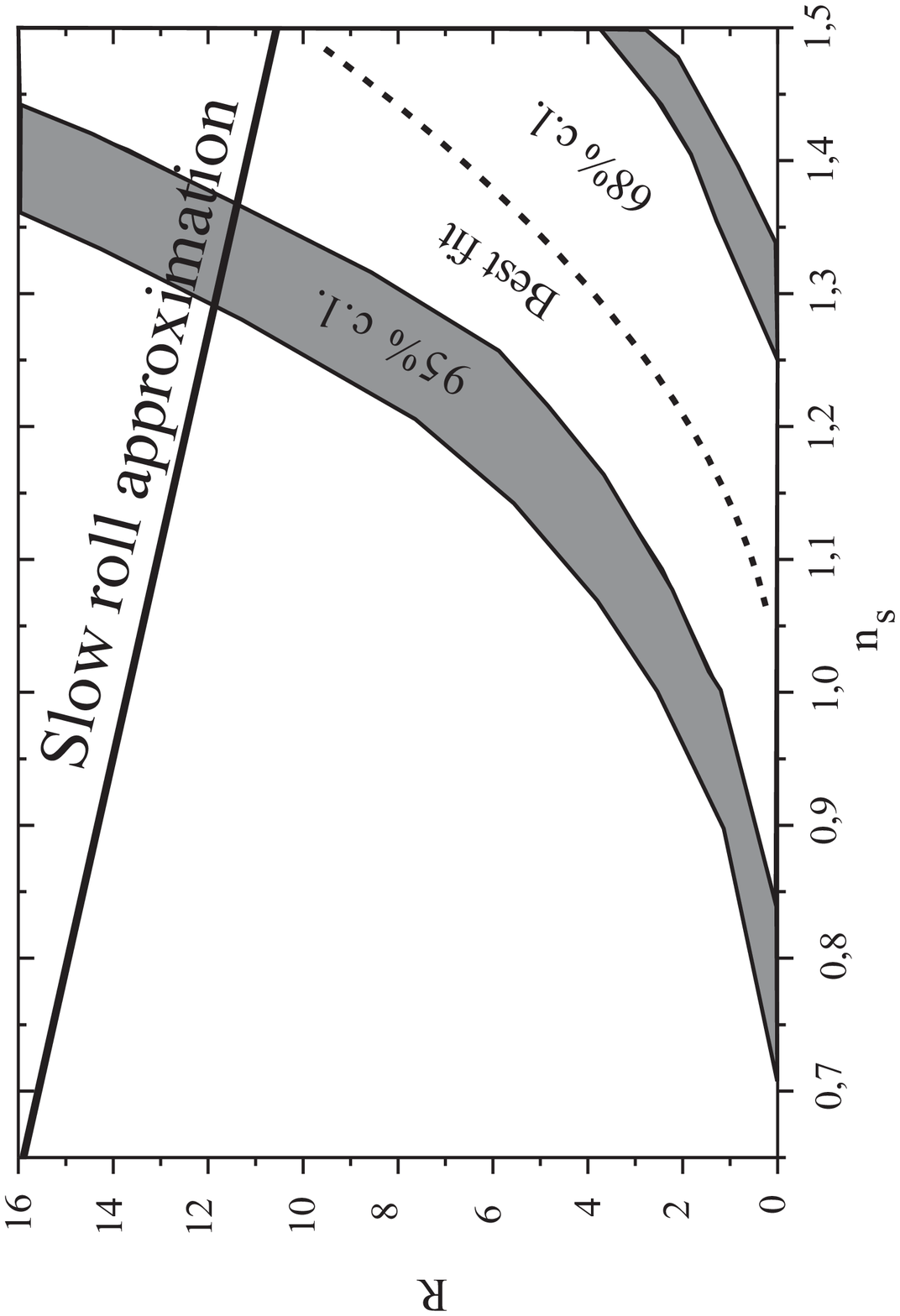}

\plotone{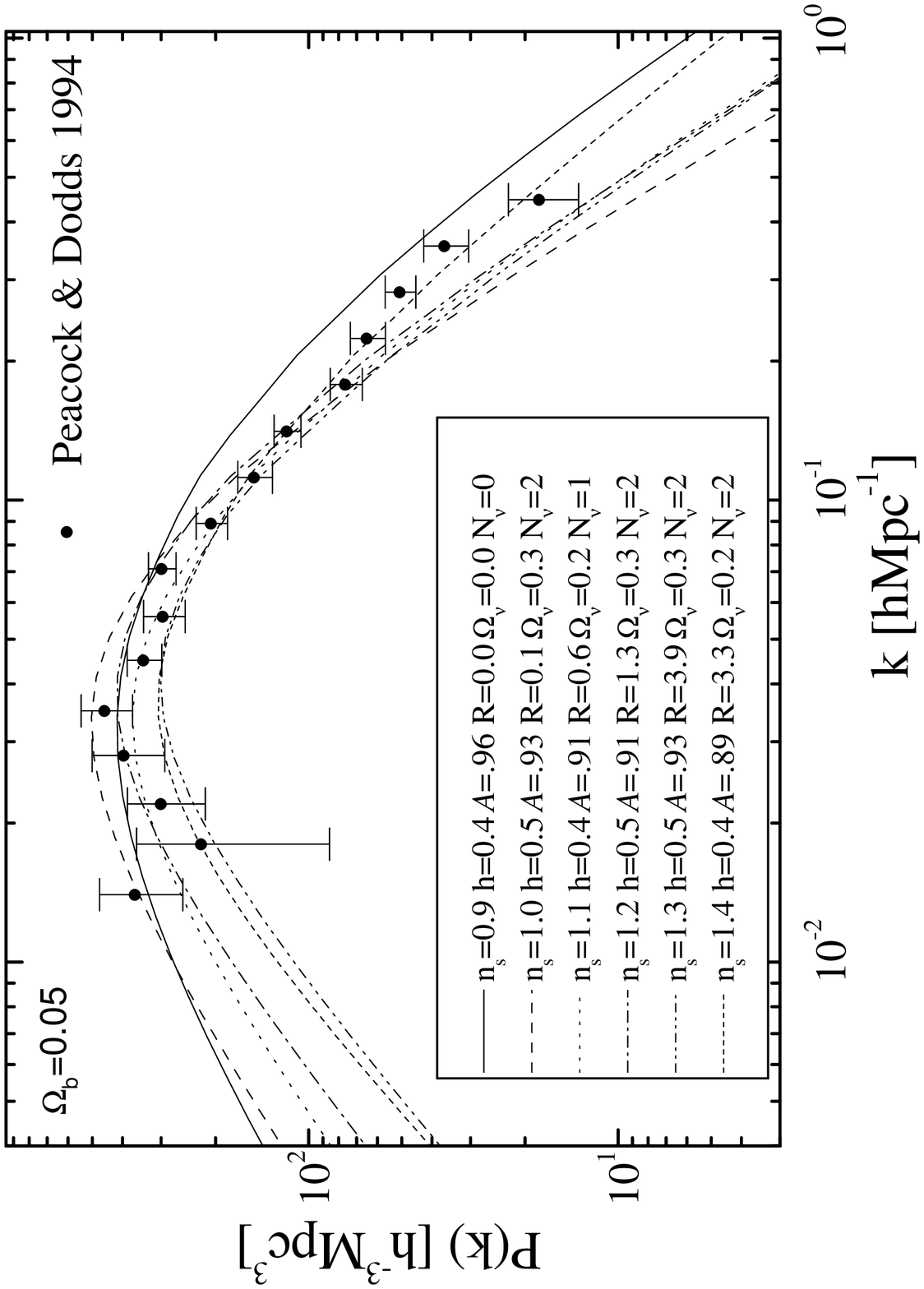}

\end{document}